\documentclass[12pt]{article}
\usepackage{amsfonts}
\usepackage{amsmath}
\usepackage{cite}

\csname @addtoreset\endcsname{equation}{section}
\newcommand{\eq}{\begin{equation}}
\newcommand{\feq}{\end{equation}}
\newcommand{\eqn}{\begin{eqnarray}}
\newcommand{\feqn}{\end{eqnarray}}
\newcommand{\arr}{\begin{eqnarray*}}
\newcommand{\farr}{\end{eqnarray*}}

\font\mybb=msbm10 at 12pt
\def\bb#1{\hbox{\mybb#1}}
\def\bZ {\bb{Z}}
\def\bR {\bb{R}}
\def\bE {\bb{E}}

\begin{document}

\begin{titlepage}
\begin{flushright}
IFUM-747-FT\\
CAMS/03-03\\
hep-th/0302218
\end{flushright}
\vspace{.3cm}
\begin{center}
\renewcommand{\thefootnote}{\fnsymbol{footnote}}
{\Large \bf Supersymmetric Domain Walls and Strings in $D=5$ gauged Supergravity coupled
to Vector Multiplets}
\vskip 25mm
{\large \bf {Sergio L.~Cacciatori$^{1,3}$\footnote{cacciatori@mi.infn.it},
Dietmar Klemm$^{2,3}$\footnote{dietmar.klemm@mi.infn.it}
and Wafic A.~Sabra$^4$\footnote{ws00@aub.edu.lb}}}\\
\renewcommand{\thefootnote}{\arabic{footnote}}
\setcounter{footnote}{0}
\vskip 10mm
{\small
$^1$ Dipartimento di Matematica dell'Universit\`a di Milano,\\
Via Saldini 50, I-20133 Milano.\\

\vspace*{0.5cm}

$^2$ Dipartimento di Fisica dell'Universit\`a di Milano,\\
Via Celoria 16, I-20133 Milano.\\

\vspace*{0.5cm}

$^3$ INFN, Sezione di Milano,\\
Via Celoria 16,
I-20133 Milano.\\

\vspace*{0.5cm}

$^4$ Center for Advanced Mathematical Sciences (CAMS) and\\
Physics Department, American University of Beirut, Lebanon.
}
\end{center}
\vspace{2cm}
\begin{center}
{\bf Abstract}
\end{center}
{\small We present new supersymmetric domain wall and string solutions
of five-dimensional $N = 2$ gauged supergravity coupled to an
arbitrary number of vector multiplets. Using the techniques of very
special geometry allows to obtain the most general domain wall
preserving half of the supersymmetries. This solution, which describes a
renormalization group flow in the dual field theory, is given in terms of
Weierstrass elliptic functions. The magnetically charged, one quarter
supersymmetric string solutions are shown to be closely related to Liouville
theory. We furthermore investigate general product space compactifications,
and show that topological transitions from AdS$_3$ $\times$ S$^2$ to
AdS$_3$ $\times$ H$^2$ can occur when one moves in moduli space.
}

\end{titlepage}

\section{Introduction}

\label{intro}

The conjectured equivalence between string theory on anti-de~Sitter (AdS)
spaces (times some compact manifold) and certain superconformal gauge
theories living on the boundary of AdS \cite{Aharony:1999ti} has led to an increasing
interest in solutions of gauged supergravities that preserve some fraction
of supersymmetry. On the CFT side, these supergravity vacua could correspond
to an expansion around non-zero vacuum expectation values of certain operators,
or describe a holographic renormalization group flow \cite{deBoer:1999xf}. In that
way, one can study strongly coupled field theories using classical supergravity
solutions. Of particular interest in this context are solutions that describe
black holes \cite{Romans:1991nq,Caldarelli:1998hg,Behrndt:1998jd}, domain
walls \cite{Behrndt:1999kz} or strings \cite{Chamseddine:1999xk,Klemm:2000nj,Maldacena:2000mw}.

In our paper we will concentrate on five-dimensional $N=2$ gauged supergravity
coupled to vector multiplets, which is relevant for holographic descriptions of
field theories in four dimensions with less than maximal supersymmetry. In particular, we
give a systematic treatment of supersymmetric domain walls and strings.
Using the tools of very special geometry, which underlies the considered supergravity
theory, allows to obtain the most general domain wall
preserving half of the supersymmetries. This solution, which describes a holographic
renormalization group flow, turns out to be given in terms of the Weierstrass elliptic
function. We show that the RG flow is a so-called vev flow, and derive the holographic
beta and ${\cal C}$-functions. The geometry has a curvature singularity signaling
nontrivial IR physics. Applying the existing criteria for allowed singularities we
obtain conditions that the invariants of the elliptic function must satisfy.

Magnetically charged string-like solutions of gauged supergravities in five dimensions
have been derived first in \cite{Chamseddine:1999xk,Klemm:2000nj}. In \cite{Klemm:2000nj}
it was suggested that solutions interpolating between AdS$_5$ and AdS$_3$ $\times$ H$^2$
have a holographic interpretation as a four-dimensional CFT that is given a
relevant perturbation and flows to a two-dimensional CFT in the IR. This aspect was
then elaborated by Maldacena and Nu\~nez \cite{Maldacena:2000mw}.
In the present work, magnetically charged strings preserving one quarter of the
supersymmetries will be studied in more detail. In particular, we show that for the
case of constant scalar fields the solutions are given in terms of functions that
satisfy the Liouville equation. They can thus be classified according to their
monodromy, which can be elliptic, hyperbolic or parabolic. The monodromy is determined
by the Liouville momentum, which turns out to be related to the curvature of the
two-dimensional Riemann surfaces into which the space transverse to the strings is
sliced.

We furthermore investigate general product space compactifications of the
five-dimensional theory. These geometries preserve half of the supersymmetries and
are of the form AdS$_3$ $\times$ S$^2$ or AdS$_3$ $\times$ H$^2$. It will be shown that
topological transitions from the former to the latter case can appear when one
moves in moduli space. In certain cases the product geometries can arise as
near-horizon limit of the one quarter supersymmetric strings mentioned above.
One has then supersymmetry enhancement near the horizon.

The remainder of this paper is organized as follows: In order to make the
paper self-contained, we briefly review $D=5$, $N=2$ gauged supergravity coupled
to vector multiplets in section \ref{sugra}. In section \ref{susystrsol} we write down
a general ansatz for supersymmetric solutions that includes both string-like
and domain wall configurations. The Killing spinor equations are analyzed and
a flow equation for the scalar fields is obtained. In section \ref{domwall} we
derive the most general half supersymmetric domain wall solution and discuss its
holographic interpretation. In the following section, the fixed points of the flow
equation are studied, and it is shown that they lead either to magnetically charged strings
with constant scalars or to product space geometries. We discuss the connection of
the magnetic strings with Liouville theory, and show the possibility of topological
transitions for the product space ``compactifications''. In section \ref{nonconst}
a new magnetic string solution with varying scalars is presented, and its
holographic interpretation is briefly discussed. Electrically charged solutions
are considered in section \ref{electrsol}.
We conclude with some final remarks in section \ref{finrem}.

\section{$D=5$, $N=2$ Gauged Supergravity}

\label{sugra}

We start with a brief description of five-dimensional $N=2$
$U(1)$-gauged supergravity theories. The fields of these theories consist of a
graviton $g_{\mu\nu}$, gravitino $\psi_{\mu}$, $n$ vector potentials
$A_{\mu}^I$ ($I=1,2,\ldots ,n$), $n-1$ gauginos $\lambda_i$ and $n-1$
scalars $\phi^i$ ($i=1,2,\ldots ,n-1$). The bosonic part of the
Lagrangian is given by \cite{Gunaydin:1984ak}

\begin{equation}
e^{-1}\mathcal{L}=\frac{1}{2}R+g^{2}V-\frac{1}{4}G_{IJ}F_{\mu \nu
}^{I}F^{J\mu \nu }-\frac{1}{2}\mathcal{G}_{ij}\partial _{\mu }\phi
^{i}\partial ^{\mu }\phi ^{j}+\frac{e^{-1}}{48}\epsilon ^{\mu \nu \rho
\sigma \lambda }C_{IJK}F_{\mu \nu }^{I}F_{\rho \sigma }^{J}A_{\lambda
}^{K}\,,  \label{action}
\end{equation}

where $\mu, \nu$ are spacetime indices, $R$ is the scalar curvature,
$F_{\mu \nu}^I$ denote the abelian field-strength tensors of the
vectors $A_{\mu}^I$, and $e = \sqrt{-g}$ is the determinant of the
f\"unfbein $e_{\mu}^a$. The scalar potential $V$ is given by

\begin{equation}
V(X) = V_I V_J \left(6X^I X^J - \frac 92 \mathcal{G}^{ij}\partial_i
X^I \partial_j X^J \right)\,,  \label{scalarpot1}
\end{equation}

where $V_I$ are constants, $\partial_i$ denotes a partial derivative
with respect to the scalar field $\phi^i$ and $X^I = X^I(\phi^i)$
are real scalars satisfying the condition $\mathcal{V} = \frac 16
C_{IJK} X^I X^J X^K = 1$. Moreover, $G_{IJ}$ and $\mathcal{G}_{ij}$ can be
expressed in terms of the homogeneous cubic polynomial $\mathcal{V}$ which
defines a ``very special geometry'' \cite{deWit:1992cr},

\begin{equation*}
G_{IJ} = -\frac 12 \frac{\partial }{\partial X^I}\frac{\partial }{\partial
X^J}\log \mathcal{V}\Big|_{\mathcal{V}=1}\,, \qquad \mathcal{G}_{ij}
= \partial_i X^I \partial_j X^J G_{IJ}\Big|_{\mathcal{V} = 1}\,.
\end{equation*}

Further useful relations can be found in appendix \ref{usefulrel}. We note
that if the five-dimensional theory is obtained by gauging a supergravity
theory coming from a Calabi-Yau compactification of M-theory, then $\mathcal{V}$
is the intersection form, $X^I$ and $X_I = \frac 16 C_{IJK} X^J X^K$
correspond to the size of the two- and four-cycles and $C_{IJK}$ are the
intersection numbers of the Calabi-Yau threefold.

The supersymmetry transformations of the gravitino $\psi_{\mu}$ and the
gauginos $\lambda_i$ in a bosonic background read \cite{Gunaydin:1984ak}

\begin{eqnarray}
\delta \psi_{\mu} &=& \left(\mathcal{D}_{\mu } + \frac i8 X_I
(\Gamma_{\mu}{}^{\nu\rho} - 4\delta_{\mu}{}^{\nu}\Gamma^{\rho})F_{\nu \rho}^I
+ \frac 12 g\Gamma_{\mu} X^I V_I \right) \epsilon \,,
\label{stgrav} \\
\delta \lambda_i &=& \left( \frac{3}{8}\Gamma ^{\mu \nu }F_{\mu \nu}^I
\partial_i X_I - \frac i2 \mathcal{G}_{ij}\Gamma ^{\mu }\partial
_{\mu }\phi^j + \frac{3i}{2}gV_I\partial_i X^I\right) \epsilon \,,
\label{stgaug}
\end{eqnarray}

where $\epsilon $ is the supersymmetry parameter and $\mathcal{D}_{\mu }$ is
the fully gauge and gravitationally covariant derivative 

\begin{equation}
\mathcal{D}_{\mu }\epsilon =\left[ \partial _{\mu }+\frac{1}{4}\omega _{\mu
ab}\Gamma ^{ab}-{\frac{3i}{2}}gV_{I}A_{\mu }^{I}\right] \epsilon .
\end{equation}

Here, $\omega_{\mu ab}$ denotes the spin connection and $\Gamma^{\mu}$ are
Dirac matrices\footnote{We use the metric $\eta _{ab} = (-,+,+,+,+)$, $\{\Gamma^a,\Gamma^b\}
= 2\eta^{ab}$, and $\Gamma^{a_{1}a_{2}\cdots a_{n}} =
\Gamma^{\lbrack {a_{1}}}\Gamma ^{a_{2}}\cdots \Gamma ^{a_{n}]}$.}.

\section{Supersymmetric String Solutions}

\label{susystrsol}

As a general ansatz for supersymmetric string-like solutions we consider
metrics of the form

\begin{equation}
ds^2 = e^{2V}(-dt^2 + dz^2) + e^{2W}(dr^2 + r^2 d\Omega_k^2)\,,
       \label{metric}
\end{equation}

where $V$ and $W$ are functions of the radial coordinate $r$ only\footnote{We apologize
for using the same symbol $V$ for the scalar potential and
for the function appearing in the metric (\ref{metric}), but the meaning should be
clear from the context.}, and
$d\Omega_k^2$ denotes the standard metric on a two-dimensional surface $\Sigma_k$
of constant scalar curvature $2k$, where $k = 0,\pm 1$. An explicit form is

\begin{equation}
d\Omega_k^2 = d\theta^2 + F_k^2(\theta) d\phi^2\,,
\end{equation}

with

\begin{equation}
F_k(\theta) = \left\{\begin{array}{l@{\,,\quad}l}
                     1 & k=0 \\ \sin\theta & k = 1 \\ \sinh\theta & k = -1\,.
                     \end{array} \right.
\end{equation}

Clearly $\Sigma_k$ is a quotient space of the universal coverings $\bE^2$ ($k=0$),
S$^2$ ($k=1$) or H$^2$ ($k=-1$).

With the choice (\ref{metric}), the f\"unfbein reads

\begin{equation}
e^0_t = e^1_z = e^V\,, \quad e^2_r = e^W\,, \quad e^3_{\theta} = e^W r\,, \quad
e^4_{\phi} = e^W r F_k\,,  
\end{equation}

and the nonvanishing components of the spin connection are given by

\begin{eqnarray}
\omega^{02}_t &=& e^{V-W}V'\,, \nonumber \\
\omega^{12}_z &=& e^{V-W}V'\,, \nonumber \\
\omega^{23}_{\theta} &=& -(W' r + 1)\,, \label{spinconn} \\
\omega^{24}_{\phi} &=& -(W' r + 1)F_k\,, \nonumber \\
\omega^{34}_{\phi} &=& -F_k'\,, \nonumber
\end{eqnarray}

where a prime denotes a derivative.

In five dimensions, strings can carry magnetic charges under the one-form
potentials $A^I$, so we assume that the gauge fields have only a magnetic part,
i.~e.

\begin{equation}
F^I_{\theta\phi} = k q^I F_k(\theta)\,, \qquad
A^I_{\phi} = k q^I \int F_k(\theta) d\theta\,. \label{magnfields}
\end{equation}

Note that $F^I$ is essentially the K\"ahler form on $\Sigma_k$.

Plugging the spin connection (\ref{spinconn}) and the magnetic fields (\ref{magnfields})
into the supersymmetry transformations of the gravitino (\ref{stgrav}), we obtain for the
Killing spinors $\epsilon$ the equations

\begin{eqnarray}
\partial_t \epsilon + \frac 12 e^{V-W}V' \Gamma_{02}\epsilon + \frac i4 k Z
e^{V-2W} r^{-2} \Gamma_{034}\epsilon + \frac 12 g e^V X^I V_I \Gamma_0\epsilon &=& 0\,,
\nonumber \\
\partial_z \epsilon + \frac 12 e^{V-W}V' \Gamma_{12}\epsilon + \frac i4 k Z
e^{V-2W} r^{-2} \Gamma_{134}\epsilon + \frac 12 g e^V X^I V_I \Gamma_1\epsilon &=& 0\,,
\nonumber \\
\partial_r \epsilon + \frac i4 k Z e^{-W} r^{-2} \Gamma_{234}\epsilon +
\frac 12 g e^W X^I V_I \Gamma_2\epsilon &=& 0\,, \nonumber \\
\partial_{\theta} \epsilon - \frac 12 (W'r + 1) \Gamma_{23}\epsilon - \frac i2
k Z e^{-W} r^{-1} \Gamma_4\epsilon + \frac 12 g e^W r X^I V_I
\Gamma_3\epsilon &=& 0\,, \nonumber \\
\partial_{\phi} \epsilon - \frac 12 (W'r + 1) F_k \Gamma_{24}\epsilon - \frac 12
F_k'\Gamma_{34}\epsilon + \frac i2 k Z e^{-W} r^{-1} F_k \Gamma_3\epsilon
&& \nonumber \\
+ \frac 12 g e^W r F_k X^I V_I \Gamma_4\epsilon - \frac{3i}{2} g k V_I q^I \int F_k
d\theta \epsilon &=& 0\,, \label{killeqns}
\end{eqnarray}

where $Z = X_I q^I$ denotes the magnetic central charge.\\
We choose as partial supersymmetry breaking conditions

\begin{equation}
\Gamma_{34}\epsilon = i\epsilon\,, \qquad \Gamma_2\epsilon = \epsilon\,.
                      \label{susybreak}
\end{equation}

This preserves one quarter of the original supersymmetries, i.~e.~, it reduces the
number of real supercharges from eight to two.

The integrability conditions following from Eqns.~(\ref{killeqns}) yield

\begin{eqnarray}
&& \partial_t \epsilon = \partial_z \epsilon = \partial_{\theta} \epsilon
= \partial_{\phi} \epsilon = 0\,, \nonumber \\
&& kZ = \frac 23 e^W r (rV' - rW' - 1)\,, \label{Z} \\
&& g X^I V_I = -\frac 13 e^{-W} (2V' + W' + r^{-1})\,, \label{gXV}
\end{eqnarray}

as well as the charge quantization condition

\begin{equation}
V_I q^I = \frac{1}{3g}\,. \label{chargequant}
\end{equation}

Notice that (\ref{chargequant}), together with $\Gamma_{34}\epsilon = i\epsilon$,
implies the twisting \cite{Maldacena:2000mw}

\begin{equation}
\omega^{34} = 3g V_I A^I\,,
\end{equation}

where $\omega^{34}$ is the spin connection on $\Sigma_k$.

Using (\ref{Z}) and (\ref{gXV}), the radial equation can be easily solved,
to give

\begin{equation}
\epsilon = e^{\frac 12 V}\epsilon_0\,, \label{killspin}
\end{equation}

where $\epsilon_0$ denotes a constant spinor subject to the constraints
(\ref{susybreak}). When the Eqns.~(\ref{Z}), (\ref{gXV}), (\ref{chargequant}) and
(\ref{killspin}) hold, the Killing spinor equations (\ref{killeqns}) are all
satisfied. What remains are the supersymmetry transformations of the gauginos
(\ref{stgaug}), which yield

\begin{equation}
\left[- e^{-2W} \frac{k}{r^2} G_{IJ} q^J + \frac 32 e^{-W} \partial_r X_I +
3gV_I\right](\partial_i X^I)\epsilon = 0\,,
\end{equation}

and thus, keeping in mind that $X_I \partial_i X^I = 0$,

\begin{equation}
- e^{-2W} \frac{k}{r^2} G_{IJ} q^J + \frac 32 e^{-W} \partial_r X_I +
3gV_I = \gamma(r) X_I\,, \label{eqgamma}
\end{equation}

where $\gamma(r)$ is some function of $r$ that can be determined by contracting
(\ref{eqgamma}) with $X^I$. In this way one obtains

\begin{equation}
\gamma(r) = - \frac 32 e^{-2W} \frac{kZ}{r^2} + 3g X^I V_I\,. \label{gamma}
\end{equation}

Using (\ref{gamma}), (\ref{Z}) and (\ref{gXV}) in (\ref{eqgamma}), we finally get
the ``flow equation'' for the scalars,

\begin{equation}
- e^{-2W} \frac{k}{r^2}G_{IJ} q^J + \frac 32 e^{-W} \partial_r X_I + 3g V_I
+ 3 e^{-W} X_I V' = 0\,. \label{flowequ}
\end{equation}

\section{The Case $k=0$: Domain Walls}

\label{domwall}

Let us first consider the case of a flat manifold $\Sigma_k$, i.~e.~$k=0$,
which will turn out to be completely integrable, with solutions given in
terms of Weierstrass elliptic functions. Below we will see that the metric
assumes then the form of a domain wall.
For $k=0$ we obtain from Eqn.~(\ref{Z})

\begin{equation}
V' = W' + r^{-1}\,, \label{VW}
\end{equation}

and hence from (\ref{gXV})

\begin{equation}
g X^I V_I = - e^{-W}(W' + r^{-1})\,. \label{gXVk=0}
\end{equation}

Using (\ref{VW}) in the flow equation (\ref{flowequ}), we get for the scalars
$X_I$

\begin{equation}
\frac 12 e^{-W} X_I' + g V_I + e^{-W} X_I (W' + r^{-1}) = 0\,, \label{flowk=0}
\end{equation}

which is easily solved to give

\begin{equation}
X_I = e^{-2W} r^{-2} [-2 g V_I  \int e^{3W} r^2 dr + C_I]\,, \label{X_IW}
\end{equation}

where the $C_I$ denote integration constants. In what follows it will be useful to
introduce the quantities

\begin{eqnarray}
a &=& C^{IJK} V_I V_J V_K\,, \qquad b = C^{IJK} V_I V_J C_K\,, \nonumber \\
c &=& C^{IJK} V_I C_J C_K\,, \qquad d = C^{IJK} C_I C_J C_K\,,
\end{eqnarray}

with $C^{IJK}$ defined in (\ref{C^IJK}).
Let us furthermore define the function

\begin{equation}
y(u) = -9a \int e^{3(W + u)} du + \frac 92 g^2 b\,,
\end{equation}

where the new radial coordinate $u$ is given by $u = \ln gr$.
Plugging (\ref{X_IW}) into Eqn.~(\ref{CXXX}), we obtain then the
differential equation

\begin{equation}
\dot y^2 = 4y^3 - g_2 y - g_3\,, \label{weierstr}
\end{equation}

where a dot denotes a derivative with respect to $u$ and

\begin{equation}
g_2 = 243 g^4 (b^2 - ac)\,, \qquad g_3 = \frac{729}{2} g^6 (3abc - a^2 d - 2b^3)\,.
\end{equation}

The general solution of Eqn.~(\ref{weierstr}) is given by

\begin{equation}
y = \wp(u + \gamma)\,,
\end{equation}

where $\wp(u)$ denotes the Weierstrass elliptic function, and $\gamma$ is an
integration constant, which we will put equal to zero without loss of generality.
$g_2$ and $g_3$ are the invariants that are related
to the periods $\omega_1$ and $\omega_2$ of the Weierstrass elliptic function by
the Eisenstein series

\begin{equation}
g_2 = 60 \sum_{m,n}{}^{\prime}\, \Omega_{m,n}^{-4}\,, \qquad
g_3 = 140 \sum_{m,n}{}^{\prime}\, \Omega_{m,n}^{-6}\,,
\end{equation}

where $\Omega_{m,n} = 2m\omega_1 + 2n\omega_2$.

Our supergravity solution is thus given by

\begin{equation}
X_I = {\dot f}^{- \frac 23} [-2g V_I f + C_I]\,,
      \label{scalweier}
\end{equation}

\begin{equation}
ds^2 = {\dot f}^{\frac 23}[- g^2 dt^2 + g^2 dz^2 + du^2 + d\Omega_0^2]\,,
       \label{metrweier}
\end{equation}

with $f(u)$ defined by

\begin{equation}
f(u) = g^{-3}\int e^{3(W + u)} du = -\frac{y(u)}{9g^3 a} + \frac{b}{2g a}\,.
       \label{f}
\end{equation}

Note that also Eqn.~(\ref{gXVk=0}), which has not been used above, is satisfied by
this solution.
From (\ref{metrweier}) we see that the metric is conformally flat like the one
of AdS$_5$, but in contrast to the latter, our solution preserves only part
of the supersymmetries, and nonconstant scalar fields are turned on.

Using the expansion

\begin{equation}
\wp(u) = u^{-2} + {\cal O}(u^2)
\end{equation}

for $u \to 0$, one sees that asymptotically the solution (\ref{metrweier})
approaches AdS$_5$ in Poincar\'{e} coordinates, and should thus have a
dual CFT interpretation. In order to investigate this
point further, we expand also the scalars,

\begin{equation}
\ln X_I - \ln X_I^0 = \frac 92 g^2 \left[\frac{C_I}{V_I} a - b\right] u^2 + {\cal O}(u^4)\,,
                      \label{expansionscal}
\end{equation}

where $X_I^0 = V_I (2/9a)^{1/3}$. Let us focus for a moment on the
STU model (which will be discussed in more detail in section \ref{nonconst}).
This model can be obtained by compactification of ten-dimensional type IIB
supergravity \cite{Cvetic:1999xp}, and has two independent scalars with mass squared
$m^2 = -4 g^2$. They are thus dual to operators of dimension
$\Delta = 2$ \cite{Aharony:1999ti}.
For this dimension, the scalars $\ln(X_I/X_I^0)$ behave as $\sim u^2 \ln u$ or $\sim u^2$
for $u \to 0$ \cite{Maldacena:2000mw}. The first kind of behaviour corresponds to the
non-normalizable mode associated to the insertion of an operator \cite{Balasubramanian:1998sn}.
From (\ref{expansionscal}) we see that in our case the operators dual to the bulk scalars
are not inserted. Instead, we have a so-called vev flow\footnote{See e.~g.~\cite{Muck:2001cy}.},
corresponding to a change of vacuum in the dual field theory, in which the operator
is given a vev.

Let us now come back to the case of an arbitrary number of vector multiplets, and
note that one can cast the solution (\ref{metrweier}) into the domain wall form

\begin{equation}
ds^2 = d\rho^2 + e^{2\omega} ds_4^2\,, \label{domainwall}
\end{equation}

where the new radial coordinate $\rho$ is defined by $d\rho = \dot{f}^{1/3} du$,
$e^{2\omega} = \dot{f}^{2/3}$, and $ds_4^2$ denotes the flat Minkowski metric in
four dimensions. (\ref{domainwall}), together with the scalars (\ref{scalweier}),
represent the most general domain wall solution to $D=5$, $N=2$ gauged supergravity
coupled to vector multiplets. From (\ref{domainwall}) we can compute the holographic
${\cal C}$-function of the renormalization group flow, given by \cite{Girardello:1998pd}

\begin{equation}
{\cal C} \propto \left(\frac{d\omega}{d\rho}\right)^{-3}\,,
\end{equation}

which yields in our case

\begin{equation}
{\cal C} \propto \frac{\dot{f}^4}{\ddot{f}^3}\,.
\end{equation}

Furthermore, if we use (\ref{gXVk=0}) as well as $e^{\omega} = r e^W$, we can rewrite
the flow equation (\ref{flowk=0}) in the form

\begin{equation}
e^{\omega}\frac{d X_I}{d e^{\omega}} = \beta_I\,, \label{Callan}
\end{equation}

where

\begin{equation}
\beta_I = \frac{2(\delta_I{}^J - X_I X^J)V_J}{X^K V_K}\,. \label{betafct}
\end{equation}

As the scalars $X_I$ represent coupling constants for the dual operators,
and the function $e^{\omega}$ determines the physical scale of the dual field
theory, we recognize (\ref{Callan}) as the Callan-Symanzik equation, with the
$I$-th beta function given by (\ref{betafct}).

Notice that for $k=0$, supersymmetry is actually enhanced from one quarter to
one half, because the constraint $\Gamma_{34}\epsilon = i\epsilon$ in
(\ref{susybreak}) can be dropped in this case. The Killing spinors for the domain
wall solution (\ref{domainwall}), (\ref{scalweier}) read

\begin{equation}
\epsilon = |\dot{f}|^{1/6}\epsilon_0\,,
\end{equation}

where the constant spinor is subject to the constraint $\Gamma_2\epsilon_0 = \epsilon_0$.

In general, the solution (\ref{metrweier}) has a curvature singularity for
$\dot f = 0$, e.~g.~the scalar curvature reads

\begin{equation}
R = \frac{4}{3\dot f^{8/3}}[\ddot f^2 - 2\dot f \dddot f]\,.
\end{equation}

In our case, the invariants $g_2$, $g_3$ are real, so that $\wp(u)$ is real
for $u \in \bR$, and one has a real and an imaginary period (although these
are not primitive periods for $\Delta \equiv g_2^3 - 27 g_3^2 < 0$). This means that
if we start from $u=0$ (AdS region) and go to positive values of $u$ (following
the RG trajectory), we will eventually reach a point $u = u_0$ where $\dot{\wp}(u)$
vanishes. The appearance of this curvature singularity is a signal of nontrivial IR
physics. According to Gubser's criterion \cite{Gubser:2000nd}, large curvatures in
geometries of the form (\ref{domainwall}) are allowed only if the scalar potential is
bounded above in the solution\footnote{It is straightforward to show that the coordinate
$\rho$ used in (\ref{domainwall}) assumes a finite value for $u \to u_0$, so Gubser's
criterion applies.}. Let us see what this restriction implies here.
To this end, we start from the potential in the form (\ref{scalarpot2}),
and use the expression (\ref{scalweier}) for the scalars. This yields

\begin{equation}
V(X) = 6\wp(u) (9a/\dot{\wp}(u))^{2/3}\,.
\end{equation}

As $(9a/\dot{\wp}(u))^{2/3}$ is always positive, and goes to $+\infty$ for
$\dot{\wp}(u) \to 0$, we must have $\wp(u_0) < 0$ in order that the potential
be bounded above in the solution. It is straightforward to show that this
implies

\begin{equation}
\Delta \equiv g_2^3 - 27 g_3^2 < 0 \qquad {\mathrm{and}} \qquad g_3 < 0\,.
\end{equation}

Apart from that, there is another physically admissible solution for $g_2 = g_3 = 0$.
In this degenerate case one has $\wp(u) = 1/u^2$, so the metric reduces to that of
AdS$_5$.

It would be interesting to lift a solution with a ``bad'' singularity to ten
dimensions. This is feasible e.~g.~for the STU model. Probably this leads to
pathologies of D3-branes with negative tension and negative charges.
We will not attempt to do this here.\\

A special case appears if the integration constants $C_I$ in (\ref{X_IW})
vanish. One has then from (\ref{weierstr}) after some simple algebra

\begin{equation}
{\dot f}^2 = \frac 23 \ddot f f\,. \label{schwarz}
\end{equation}

In other words, the Schwarzian derivative $(2\ddot f f - 3{\dot f}^2)/2f^2$
vanishes. Using $f = \int e^{3W} r^2 dr$ following from (\ref{f}),
the differential equation (\ref{schwarz}) yields

\begin{equation}
e^W r = C (W'r + 1)\,, \label{1stint}
\end{equation}

$C$ denoting an integration constant. (\ref{1stint}) is actually a first
integral of the Liouville equation with zero ``momentum''. To see this, consider
the Liouville equation

\begin{equation}
\Delta W = \mu e^{2W}\,, \label{Liouvgen}
\end{equation}

where $\mu > 0$ is a constant, in spherical coordinates $(r, \sigma)$,
for Liouville fields $W$ independent of the angular coordinate $\sigma$.
In this case one has from (\ref{Liouvgen})

\begin{equation}
W'' + \frac{W'}{r} = \mu e^{2W}\,. \label{Liouvr}
\end{equation}

The action from which (\ref{Liouvr}) follows is invariant under the transformation
$r \to \lambda r$, $W \to W - 2\ln \lambda$, with $\lambda \in \bR^+$. The
first integral associated to this invariance is

\begin{equation}
(W' r + 1)^2 = \mu r^2 e^{2W} - p^2\,, \label{intLiou}
\end{equation}

where the integration constant $p$ is essentially the Liouville
momentum \cite{Seiberg:1990eb,Ginsparg:is}. Comparing with (\ref{1stint}),
we see that $W$ has zero momentum, and corresponds thus to a parabolic
Liouville solution \cite{Seiberg:1990eb}, given by

\begin{equation}
e^W = \frac{C}{r\ln gr}\,. \label{parabol}
\end{equation}

One can now use (\ref{parabol}) in (\ref{X_IW}) to determine the scalar fields,
with the result

\begin{equation}
X_I = g C V_I\,,
\end{equation}

and thus the scalars are forced to be constants in the case $k=0$, $C_I = 0$.

Determining finally $V$ from (\ref{VW}) yields the metric

\begin{equation}
ds^2 = \frac{C^2}{\ln^2 gr}[-dt^2 + dz^2 + \frac{dr^2}{r^2} + d\Omega_0^2]\,.
       \label{AdS5}
\end{equation}

Introducing the coordinate $u = \ln gr$, one easily sees that (\ref{AdS5})
is AdS$_5$ in Poincar\'{e} coordinates, and thus preserves all supersymmetries.

\section{Fixed Points of the Flow Equation}

\label{fixedpts}

We now come back to the case of general $k$, and would like to
determine the fixed points of the flow equation (\ref{flowequ}), i.~e.~, the
points where $\partial_r X_I = 0$, and thus

\begin{equation}
- e^{-2W} \frac{k}{r^2}G_{IJ} q^J + 3g V_I + 3 e^{-W} X_I V' = 0\,. \label{flowfix}
\end{equation}

From (\ref{Z}) and (\ref{gXV}) we get

\begin{equation}
V' = \frac{1}{2r^2} kZ e^{-W} - g X^I V_I e^W\,. \label{Vprime}
\end{equation}

Inserting this into (\ref{flowfix}) yields

\begin{equation}
e^{-2W} \frac{k}{r^2}(-G_{IJ}q^J + \frac 32 Z X_I) + 3g (V_I - X_I X^J V_J) = 0\,.
\end{equation}

Obviously we have to distinguish two cases, namely $e^W r$ constant and $e^W r$ not
constant. Let us first study the latter case. From the foregoing equation we have
then

\begin{equation}
X^I = \frac{q^I}{Z}\,, \qquad X_I = \frac{V_I}{X^J V_J}\,.
\end{equation}

The values $X^I = q^I/Z$ for the scalars are exactly the ones which extremize the
magnetic central charge $Z$\footnote{This is analogous to the extremal electric central
charge for black holes in the ungauged theory \cite{Ferrara:1996dd}.}.
The critical value of $Z$ reads

\begin{equation}
Z = \left(\frac 16 C_{IJK} q^I q^J q^K\right)^{\frac 13}\,.
\end{equation}

Using $q^I = Z X^I$ in the charge quantization condition (\ref{chargequant}),
one obtains

\begin{equation}
X^I V_I = \frac{1}{3gZ}\,. \label{XV}
\end{equation}

We will now show that similar to the case $k=0$, the Liouville equation appears
also for $k = \pm 1$. To see this, define the field $\Phi$ by

\begin{equation}
e^{2\Phi} = e^{2W} + \frac{3k Z^2}{r^2}\,.
\end{equation}

Making use of (\ref{Z}), (\ref{gXV}) and (\ref{XV}), it is straightforward to
show that

\begin{equation}
(\Phi' r + 1)^2 = \frac{r^2}{9Z^2} e^{2\Phi} - \frac k3\,,
\end{equation}

which is again a first integral of the Liouville equation, with momentum
given by $p^2 = k/3$, as can be seen by comparing with (\ref{intLiou}).
The field $\Phi$ satisfies thus the Liouville equation

\begin{equation}
\Phi'' + \frac{\Phi'}{r} = \frac{1}{9Z^2} e^{2\Phi}\,.
\end{equation}

For $k=1$, we have real momentum, so the solution is hyperbolic \cite{Seiberg:1990eb},

\begin{equation}
e^{2\Phi} = \frac{3Z^2}{r^2\sin^2 \left(\frac{\ln gr}{\sqrt 3}\right)}\,,
            \label{hyperbol}
\end{equation}

whereas for $k=-1$ (imaginary momentum), the solution is elliptic,

\begin{equation}
e^{2\Phi} = \frac{3Z^2}{r^2\sinh^2 \left(\frac{\ln gr}{\sqrt 3}\right)}\,.
            \label{ellipt}
\end{equation}

Determining also $V$ from (\ref{Vprime}), one obtains finally for the metric

\begin{equation}
ds^2 = \frac{k}{\cos(p\ln gr)\sin^2(p\ln gr)} (-dt^2 + dz^2) +
       \frac{3kZ^2}{r^2}\cot^2(p\ln gr)(dr^2 + r^2 d\Omega_k^2)\,, \label{finalmetr}
\end{equation}

with the Liouville momentum $p = \sqrt{k/3}$. (\ref{finalmetr}) coincides with the
solutions found in \cite{Chamseddine:1999xk,Klemm:2000nj}, which are written here
in different coordinates that make the connection with Liouville theory more
evident\footnote{It is worth pointing out that also the magnetic branes of
Einstein-Maxwell-AdS gravity in arbitrary dimension considered
in \cite{Sabra:2002xy} can be written in terms of functions satisfying
the Liouville equation.}.
We observe that the line element (\ref{finalmetr}) is invariant under the transformations
$gr \to 1/gr$ or $gr \to gr\exp(2\pi n/p)$, where $n \in \bZ$. In terms of the coordinate
$u = \ln gr$ this means $u \to -u$ or $u \to u + 2\pi n/p$. Note that $p$ is real
only for $k=1$. In this case, one encounters a naked curvature singularity for
$pu = \pi/2$. For $k=-1$, the singularity is hidden by a horizon, which appears
for $u \to \infty$. The solution approaches AdS$_3$ $\times$ H$^2$ near the horizon.
It will be shown below that this geometry has enhanced supersymmetry.
Notice finally that the conformal boundary of the solution (\ref{finalmetr}) is
reached for $u \to 0$.\\

We come now to the case where $e^W r$ is constant. From (\ref{metric})
one sees that this corresponds to a product space $M_3 \times \Sigma_k$,
with $M_3$ a three-manifold to be determined below. Let us decompose
$q^I$ and $V_I$ in a part parallel to $X^I$ and a part orthogonal to $X^I$,

\begin{eqnarray}
q^I &=& Z X^I + Z P^I{}_J a^J\,, \label{decompq} \\
V_I &=& X^J V_J X_I + c_J P^J{}_I\,, \label{decompV}
\end{eqnarray}

where $P^I{}_J = \delta^I{}_J - X^I X_J$ is a projector satisfying
$P^I{}_J X^J = X_I P^I{}_J = 0$, and $a^I$, $c_I$ parametrize the parts
orthogonal to $X^I$. Plugging the above decompositions into (\ref{flowfix})
and eliminating $V'$ by means of (\ref{Z}) yields

\begin{eqnarray}
kZ &=& - g r_0^2 X^I V_I\,, \label{Zr0} \\
d_I &=& \frac{kZ}{3 g r_0^2} G_{IJ} b^J\,, \label{dI}
\end{eqnarray}

where we defined $r_0 = e^W r$, $b^I = P^I{}_J a^J$ and $d_I = c_J P^J{}_I$.
Using Eqns.~(\ref{decompq}), (\ref{decompV}) and (\ref{dI}) in (\ref{Zr0}),
one gets

\begin{equation}
r_0^2 = k Z^2 [G_{IJ} b^I b^J - 3]\,. \label{r02}
\end{equation}

From (\ref{r02}) we conclude that $k=1$ for $G_{IJ} b^I b^J > 3$, and $k=-1$
for $G_{IJ} b^I b^J < 3$. For $G_{IJ} b^I b^J = 3$ an interesting topological
transition occurs\footnote{In \cite{Klemm:2000nj} only the case $b^I = 0$ was
considered. This leads to $k=-1$.}.
Start e.~g.~with positive $G_{IJ} b^I b^J - 3$, so that
$k=1$, i.~e.~$\Sigma_k$ is a two-sphere. Let then $G_{IJ} b^I b^J - 3$ go
to zero, which means that the radius $r_0$ goes to zero, so that the S$^2$ shrinks
to a point. When $G_{IJ} b^I b^J - 3$ changes sign to become negative, the
two-manifold $\Sigma_k$ restarts to blow up, but now as a hyperbolic space H$^2$.
These topological transitions are similar to the ones considered
in \cite{Aspinwall:1993yb}\footnote{Note however that in
\cite{Aspinwall:1993yb} cycles in the internal Calabi-Yau manifold
shrink to zero and then blow up into a different Calabi-Yau, whereas in our case
the moduli $X^I$ do not go to zero at the transition.}.
Let us examine in more detail the transition point. Using (\ref{decompq}) and
(\ref{G_IJ}), one easily shows that $G_{IJ} b^I b^J = 3$ is equivalent to

\begin{equation}
C_{IJK} q^I q^J X^K = 0\,.
\end{equation}

It would be interesting to understand the meaning of this topological transition
in the dual conformal field theory.

We still have to determine the function $V$. Integrating Eqn.~(\ref{Z})
one obtains

\begin{equation}
e^{2V} = (gr)^{3kZ/r_0}\,.
\end{equation}

If we finally introduce the new coordinate $\rho$ defined by
$(g\rho)^2 = (gr)^{3kZ/r_0}$, the five-dimensional metric reads

\begin{equation}
ds^2 = (g\rho)^2(-dt^2 + dz^2) + \left(\frac{2r_0^2}{3kZ}\right)^2
       \frac{d\rho^2}{\rho^2} + r_0^2 d\Omega_k^2\,,
\end{equation}

so that the manifold is AdS$_3$ $\times \Sigma_k$. 

It turns out that in the case of constant $e^W r$, the constraint
$\Gamma_2 \epsilon = \epsilon$ on the Killing spinors can be dropped,
so that we only impose $\Gamma_{34}\epsilon = i\epsilon$. The above
product space solutions preserve thus half of the supersymmetry.
The Killing spinors read

\begin{equation}
\epsilon = e^{V/2} P \left[1 + \frac{3kZ}{2r_0^2}(\Gamma_0 t + \Gamma_1 z)\right]
           \Pi \epsilon_0 + e^{-V/2} (1 - P) \Pi \epsilon_0\,,
\end{equation}

where we defined the projectors $P = (1 + \Gamma_2)/2$, $\Pi = (1 - i\Gamma_{34})/2$,
and $\epsilon_0$ denotes an arbitrary constant spinor.

\section{Nonconstant Scalars and $k=\pm 1$}

\label{nonconst}

We now consider the flow equation (\ref{flowequ}) for the case of
nonconstant scalar fields and $k=\pm 1$. To be specific, we consider the STU model
which has only one intersection number $C_{123} = 1$ nonzero. This model
can be embedded into gauged $N=4$ and $N=8$ supergravity as well.
Furthermore, it can be obtained by compactification of ten-dimensional type
IIB supergravity \cite{Cvetic:1999xp}. This allows to lift the solutions
presented below to ten dimensions.

The prepotential reads

\begin{equation}
{\cal V} = STU = 1\,.
\end{equation}

Taking $S= X^1$, $T = X^2$ and $U = X^3$ one gets for the matrix $G^{IJ}$

\begin{equation}
G^{IJ} = 2\left(\begin{array}{ccc} S^2 & & \\ & T^2 & \\ & & U^2
                \end{array} \right)\,.
\end{equation}

Without loss of generality we assume $V_I = 1/3$.
Considering $S$ as the dependent field (i.~e.~$S = 1/TU$) we find for the
potential

\begin{equation}
V(X) = 2(\frac 1U + \frac 1T + TU)\,,
\end{equation}

which has a minimum $V_{min}(U = T = 1) = 6$.

The equations to solve are given in appendix \ref{solnonconst}.
Maldacena and Nu\~nez \cite{Maldacena:2000mw} found solutions to these equations with
nonconstant scalars for the special case $T=U$ and $q^2 = q^3 = 0$,
where the operator dual to $\ln T$ is inserted. We will present here a different solution,
where the operators dual to the bulk scalars are not inserted, but instead
they are given a vev by a change of vacuum in the dual field theory.

In appendix \ref{solnonconst} it is shown that for the STU model, a special solution
to the Eqns.~(\ref{Z}), (\ref{gXV}), (\ref{chargequant}) and (\ref{flowequ}) is
given by

\begin{eqnarray}
&& q^1 = q^2 = q^3 = \frac 1{3g}\,, \nonumber \\
&& ds^2 = e^{\frac k6 u^2} u^{-2} \left[1 - \frac k6 u^2\right]^{-2/3}
          (-dt^2 + dz^2) + g^{-2} u^{-2}\left[1 - \frac k6 u^2\right]^{4/3}
          (du^2 + d\Omega_k^2)\,, \nonumber \\
&& S = U = \left[1 - \frac k6 u^2\right]^{1/3}\,, \label{sergio} \\
&& T = S^{-2}\,, \nonumber
\end{eqnarray}

where again $u = \ln gr$. The conformal boundary is approached for $u \to 0$.
As in (\ref{expansionscal}) we can expand the moduli for $u \to 0$, and see that
they behave as the normalizable mode, so that the dual operators are not inserted.
Note that if $k=1$, the above metric becomes singular for $u^2 = 6$. In order to
decide whether this singularity is allowed or not, one cannot use the criterion
of \cite{Gubser:2000nd}, because strictly speaking this applies only to geometries
of the domain wall form. We can however lift our solution to ten dimensions using
the rules given in \cite{Cvetic:1999xp}, and then use the criterion of \cite{Maldacena:2000mw}
which states that the component $g_{00}$ of the ten-dimensional metric should not
increase as we approach the singularity. In our case it is easy to show that
$g_{00}$ blows up like $(1 - u^2/6)^{-1}$, so that the singularity would not be
allowed according to \cite{Maldacena:2000mw}.

\section{Electric Solutions}

\label{electrsol}

One can try to find an electric analogue of (\ref{metric}). An
obvious ansatz would be

\begin{equation}
ds^2 = e^{2V}(dx^2 + dy^2) + e^{2W}(dr^2 + r^2 d\sigma_k^2)\,,
\end{equation}

where again $V$ and $W$ are functions of $r$ only, and $d\sigma_k^2$
denotes the standard metric on two-dimensional Minkowski space ($k=0$),
de~Sitter space dS$_2$ ($k=1$) or anti-de~Sitter space AdS$_2$ ($k=-1$).
A possible choice is

\begin{equation}
d\sigma_k^2 = -dt^2 + f_k^2(t) dz^2\,,
\end{equation}

with

\begin{equation}
f_k(\theta) = \left\{\begin{array}{l@{\,,\quad}l}
                     1 & k=0 \\ \sinh t & k = 1 \\ \sin t & k = -1\,.
                     \end{array} \right.
\end{equation}

For the gauge fields we take

\begin{equation}
F^I_{tz} = k q^I f_k(t)\,, \qquad
A^I_z = k q^I \int f_k(t) dt\,. \label{elfields}
\end{equation}

Using projection conditions on the Killing spinors analogous to
(\ref{susybreak}), one finds that the charge quantization condition
is now given by

\begin{equation}
V_I q^I = \frac{1}{3ig}\,,
\end{equation}

so that either the charges $q^I$ or the coupling constant $g$ have to be
imaginary, which is of course unphysical.
Nevertheless the possibility of having imaginary coupling constant $g$
suggests that supersymmetric electric solutions of the type considered here
might exist in the exotic de~Sitter supergravity theories introduced by
Hull \cite{Hull:1998vg}.

\section{Final Remarks}

\label{finrem}

We conclude this paper by pointing out some possible extensions of the
work presented here. First of all, one notes that for the magnetic strings,
the solutions with parabolic, hyperbolic or elliptic
monodromy (cf.~Eqns.~(\ref{parabol}), (\ref{hyperbol}) and (\ref{ellipt}))
essentially coincide with the Weierstrass elliptic
function $\wp(u)$ (with $u = \ln gr$) for the degenerate case where the discriminant
$\Delta = g_2^3 - 27 g_3^2$ vanishes (see e.~g.~\cite{akhiezer}). This suggests
that there might exist more general magnetically charged string solutions, whose
metric is given in terms of elliptic functions. If these solutions preserve some
supersymmetry, it is clear from the results of our paper that nonconstant scalars
have to be turned on in this case.

Furthermore, it would be interesting to find the nonextremal or rotating
generalizations\footnote{In four dimensions, the rotating generalization of the
magnetically charged, one quarter supersymmetric soliton was found
in \cite{Caldarelli:1998hg}.} of the supersymmetric strings found here.
Holographically, this would correspond to considering respectively field theories
at finite temperature or on a rotating manifold.

We hope to report on these points in a future publication.

\section*{Acknowledgements}
\small

This work was partially supported by INFN, MURST and
by the European Commission RTN program
HPRN-CT-2000-00131, in which S.~L.~C.~and D.~K.~are
associated to the University of Torino.
The authors would like to thank J.~Maldacena and W.~M\"uck for clarifying
correspondence.
\normalsize

\newpage

\begin{appendix}

\section{Useful Relations in Very Special Geometry}

\label{usefulrel}

We list here some useful relations that can be proven using the
techniques of very special geometry:

\begin{equation}
\partial_i X_I = -\frac 23 G_{IJ} \partial_i X^J\,, \qquad
X_I = \frac 23 G_{IJ} X^J\,.
\end{equation}

\begin{equation}
G_{IJ} = \frac 92 X_I X_J - \frac 12 C_{IJK} X^K\,, \label{G_IJ}
\end{equation}

\begin{equation}
G^{IJ} = 2 X^I X^J - 6 C^{IJK} X_K\,, \label{G^IJ}
\end{equation}

where the $C^{IJK}$ are defined by

\begin{equation}
C^{IJK} = \delta^{II'}\delta^{JJ'}\delta^{KK'} C_{I'J'K'}\,. \label{C^IJK}
\end{equation}

Eqn.~(\ref{G^IJ}) can be shown using the ``adjoint identity''

\begin{equation}
C_{IJK} C_{J'\left(LM\right.} C_{\left.PQ\right)K'} \delta^{JJ'}\delta^{KK'}
= \frac 43 \delta_{I\left(L\right.} C_{\left.MPQ\right)}
\end{equation}

of the associated Jordan algebra \cite{Gunaydin:1983bi}. Using (\ref{G^IJ})
one obtains furthermore

\begin{equation}
X^I = \frac 92 C^{IJK} X_J X_K\,,
\end{equation}

and

\begin{equation}
\frac 92 C^{IJK} X_I X_J X_K = 1\,. \label{CXXX}
\end{equation}

The scalar potential (\ref{scalarpot1}) can also be written as \cite{Behrndt:1998jd}

\begin{equation}
V(X) = 9 V_I V_J \left(X^I X^J - \frac 12 G^{IJ}\right)\,.
\end{equation}

Using (\ref{G^IJ}), this can be cast into the form

\begin{equation}
V(X) = 27 C^{IJK} V_I V_J X_K\,. \label{scalarpot2}
\end{equation}

\section{Solutions with nonconstant Scalars for $k = \pm 1$}
\label{solnonconst}

For the STU model considered in section \ref{nonconst}, we
introduce the rescaled fields $x^I = re^W X^I$ so that $x^1 x^2 x^3 = r^3 e^{3W}$.
The equations (\ref{chargequant}), (\ref{gXV}), (\ref{Z}) and (\ref{flowequ}) become

\eqn
& & g \sum_I q^I = 1\,, \\
& & g \sum_I x^I = -\left( 2\dot V +\dot W +1 \right)\,, \label{11} \\
& & k \sum_I \frac{q^I}{x^I} = -2 \left( \dot W +1-\dot V \right) \label{22}\,, \\
& & -kq^I -\dot x^I +2gx^{I2} +(2\dot V +\dot W +1 )x^I = 0\,, \label{33}
\feqn

where the dot denotes a derivative with respect to $u = \ln gr$.
Using (\ref{11}) in (\ref{33}) we find

\eqn
& & kq^1 +\dot x^1 + g(-x^1 + x^2 + x^3) x^1 = 0\,, \nonumber \\
& & kq^2 +\dot x^2 + g(x^1 - x^2 + x^3) x^2 = 0\,, \label{system} \\
& & kq^3 +\dot x^3 + g(x^1 + x^2 - x^3) x^3 = 0\,. \nonumber
\feqn

Note that from $x^1 x^2 x^3 = r^3 e^{3W}$ we derive

\eqn
3(\dot W + 1) = \sum_I \frac{\dot{x}^I}{x^I}\,,
\feqn 

which confronted with the sum of (\ref{11}) and (\ref{22}) gives the 
consistency condition

\eqn
\sum_I \frac{\dot{x}^I}{x^I} = -g \sum_I x^I - k \sum_I \frac{q^I}{x^I}\,,
\feqn

which is satisfied by the system (\ref{system}).

One can then solve the Eqns.~(\ref{system}) and use
$x^1 x^2 x^3 = r^3 e^{3W}$ to find $W$ and finally (\ref{11}) or (\ref{22})
to determine $V$ so that

\eqn
ds^2 = (\prod_I x^I)^{-\frac 13} e^{-g \int \sum_J x^J du}
(-dt^2 +dz^2) +(\prod_I x^I)^{\frac 23} (du^2 + d\Omega_k^2 )\,.
\feqn

If we introduce the functions

\eqn
& & y^1 = x^1 + x^2 - x^3\,, \nonumber \\
& & y^2 = x^1 - x^2 - x^3\,, \nonumber \\
& & y^3 = x^1 - x^2 + x^3\,, \nonumber
\feqn

and the constants

\eqn
& & Q^1 = -k(q^1 +q^2 -q^3 )\,, \nonumber \\
& & Q^2 = -k(q^1 -q^2 -q^3 )\,, \nonumber \\
& & Q^3 = -k(q^1 -q^2 +q^3 )\,, \nonumber
\feqn

the system (\ref{system}) becomes

\eqn
& & Q^1 - \dot y^1 + g y^2 y^3 = 0\,, \label{y1} \\
& & Q^2 - \dot y^2 + g y^1 y^3 = 0\,, \label{y2} \\
& & Q^3 - \dot y^3 + g y^1 y^2 = 0\,. \label{y3}
\feqn

From (\ref{y1}) and (\ref{y2}) one finds

\eqn
& & y^1 + y^2 = e^{g \int y^3 du} \left[(Q^1 + Q^2) \int e^{- g\int y^3 du} du + K_+
\right]\,, \nonumber \\
& & y^1 - y^2 = e^{-g \int y^3 du} \left[(Q^1 - Q^2) \int e^{g\int y^3 du} du + K_- \right]\,,
          \nonumber
\feqn

where $K_{\pm}$ denote integration constants.
Using these expressions in (\ref{y3}) one obtains an integro-differential equation for
$y^3$ which is quite complicated. However a solution can be found e.~g.~in the simple
case $Q^1 + Q^2 = 0$ and $K_+ = 0$. This corresponds to

\begin{displaymath}
q^1 = q^3\,, \qquad q^2 = \frac 1g - 2q^1\,.
\end{displaymath}

If we introduce the function

\begin{displaymath}
h = Q^1 \int e^{g\int y^3 du} du + \frac{K_-}{2}\,,
\end{displaymath}

we obtain

\begin{displaymath}
y^1 = -y^2 = Q^1\frac{h}{\dot h}\,.
\end{displaymath}

Eqn.~(\ref{y3}) becomes

\eqn
\partial^2_u \ln\dot h = g Q^3 - \left(\frac{g Q^1 h}{\dot h}\right)^2\,.
                              \label{diffequh}
\feqn

A special solution can be found using the ansatz

\eqn
h(u) = h_0 e^{\alpha u^2}\,,
\feqn

which solves (\ref{diffequh}) if

\[
\alpha = \frac g2 Q^3\,, \qquad Q^1 = \pm Q^3\,.
\]

Choosing the plus\footnote{The minus sign leads to a solution where some of
the moduli $X^I$ become negative. We discard this unphysical case.} we get
the solution (\ref{sergio}).

\end{appendix}

\end{document}